\documentclass[12pt,preprint]{emulateapj}

\def\mnras{{ MNRAS}}

\def\be{\begin{equation}}
\def\ee{\end{equation}}
\def\bea{\begin{eqnarray}}
\def\eea{\end{eqnarray}}

\usepackage{color,txfonts}

\usepackage{graphicx}

\begin{document}
\title{Implications of the Tentative Association between GW150914 and a {\it Fermi}-GBM Transient}
%\title{Implication of the association between GBM transient 150914 and LIGO Gravitational Wave event GW150914}
\author{Xiang Li$^{1,3}$, Fu-Wen Zhang$^{2,1}$, Qiang Yuan$^{1}$, Zhi-Ping Jin$^{1}$,  Yi-Zhong Fan$^{1}$, Si-Ming Liu$^{1}$, and Da-Ming Wei$^{1}$}
\affil{
$^1$ {Key Laboratory of dark Matter and Space Astronomy, Purple Mountain Observatory, Chinese Academy of Science, Nanjing, 210008, China.}\\
$^2$ {College of Science, Guilin University of Technology, Guilin 541004, China.}\\
$^3$ {University of Chinese Academy of Sciences, Yuquan Road 19, Beijing, 100049, China.\\}
}
\email{fwzhang@glut.edu.cn (FWZ), yzfan@pmo.ac.cn (YZF), dmwei@pmo.ac.cn (DMW)}

\begin{abstract}
The merger-driven Gamma-ray Bursts (GRBs) and their associated gravitational wave (GW) radiation, if both successfully detected, have some far-reaching implications, including for instance: (i) The statistical comparison of the physical properties of the short/long-short GRBs with and without GW detection can test the general origin model; (ii) Revealing the physical processes taking place at the central engine; (iii) Measuring the velocity of the Gravitational wave directly/accurately. In this work we discuss these implications in the case of possible association of GW150914/ GBM transient 150914. We compared GBM transient 150914 with other SGRBs and found that such an event {may be} a distinct outlier in some statistical diagrams, possibly due to its specific binary-black-hole merger origin. However, the presence of a ``new" group of SGRBs with ``unusual" physical parameters  is also possible. If the outflow of GBM transient 150914 was launched by the accretion onto the nascent black hole, the magnetic activity rather than the neutrino process is likely responsible for the energy extraction and the accretion disk mass is estimated to be $\sim 10^{-5}~M_\odot$. The GW150914/GBM transient 150914 association, {if confirmed, would} provide the first opportunity to directly measure the GW velocity and its departure  from the speed of the light {should be within} a factor of $\sim 10^{-17}$.
\end{abstract}

\section{Introduction}
The mergers of compact object binaries are known to be promising gravitational wave sources and are prime targets of advanced LIGO/Virgo network \citep[e.g.,][]{Clark1977,Aasi2013}. Such mergers involving neutron stars are also widely believed to be the physical origin of SGRBs \citep[e.g.,][]{Eichler1989,Piran2004,Kumar2015} that lasted typically shorter than 2 seconds in soft $\gamma-$ray band
\citep{Kouveliotou1993}. After the discovery of the so-called long-short events GRB 060505 and in particular GRB 060614 (both are apparently long-lasting but do not show any signal of supernovae down to very stringent limits; see \citet{Fynbo2006}), it had been suspected that the compact object mergers could produce these peculiar events as well \citep{Gehrels2006,DellaValle2006,Gal-Yam2006,Zhang2007}. Before Sept. 2015, due to the lack of direct detection of gravitational wave (GW),  the evidence for the compact object merger origin of SGRBs are from the observations of their afterglows and host galaxies \citep{Berger2014}. The most important indirect evidence may be the identification of Li-Paczy\'{n}ski macronovae/kilonovae \citep[e.g.,][]{Li1998,Metzger2010,Barnes2013}, arising from the radioactive decay of $r-$process material synthesized in the ejecta that is launched during the mergers, in SGRB 130603B \citep{Tanvir2013,Berger2013}, long-short GRB 060614 \citep[lsGRB 060614;][]{Yang2015,Jin2015} and SGRB 050709 \citep{Jin2016}, which in turn suggests that compact object mergers do take place. Interestingly, the macronova/kilonova modeling of the signals in lsGRB 060614 and SGRB 050709 favors the mergers of neutron star-black hole binaries. The expected advanced-LIGO/VIRGO sensitivity range for neutron star-black hole merger events is about twice that of the binary neutron star merger events \citep{Aasi2013}. Benefitted from such an ``improvement", lsGRB 060614 and possibly also lsGRB 060505 are within the expected advanced-LIGO/VIRGO sensitivity range \citep{LiX2016}. Such a finding is very encouraging for the people interested in establishing GRB/GW association since no known SGRB has been found within the advanced-LIGO/VIRGO sensitivity range for binary neutron star system \citep[e.g.][]{Clark2015}. The detection rate of GRB/GW association by the advanced LIGO/VIRGO network in its full performance is estimated to be ${\cal R}_{\rm GRB/GW}\sim 1~{\rm yr}^{-1}$ and that GRB/GW association is widely expected to be not formally established until 2020.

On September 14, 2015 at 09:50:45 UTC the two detectors of the Laser Interferometer Gravitational-Wave
Observatory (i.e., LIGO) simultaneously detected a transient gravitational-wave signal sweeping upwards in
frequency from 35 to 250 Hz with a peak gravitational-wave strain of $1.0\times 10^{-21}$ and matching the waveform
predicted by general relativity for the inspiral and merger of a pair of $\sim 30 M_\odot$ black holes and the ringdown of the
single newly-formed massive black hole \citep{Abbott2016}. This great event is known as GW 150914, which is the first direct
detection of gravitational waves and the first identification of a binary black hole merger \citep{Abbott2016}. {\it Surprisingly,}
the Fermi Gamma-ray Burst Monitor (GBM) observations at the time of GW150914 reveal the presence of a weak
gamma-ray transient 0.4 s after the gravitational wave event was recorded (i.e., the delay between the GW signal and the GRB onset is $\delta t\sim 0.4$ s), with a false alarm probability of 0.0022 \citep{Connaughton2016}. This weak but hard gamma-ray transient lasted $T_{\gamma} \sim 1$ s and its localization, though poorly-constrained, is consistent with that of GW150914. With the luminosity distance $D\sim 410$ Mpc of GW150914, the isotropic-equivalent energy of the gamma-ray transient released between 1 keV and 10 MeV is of $L_{\rm \gamma}=1.8^{+1.5}_{-1.0}\times 10^{49}~{\rm erg~s^{-1}}$, which is also typical for SGRBs  \citep{Connaughton2016}.
Nevertheless, we call the possible $\gamma-$ray event as ``GBM transient 150914" rather than ``SGRB 150914" because the simultaneous observations by INTEGRAL  \citep{Savchenko2016} did not yield a similar signal (See \citet{Connaughton2016} for possible solution of the tension between these observation results). In this work we focus on the implications of the association between GW150914 and the possible GBM transient 150914.

%This work is structured as the following. In Sec.2 we introduce some implications of the GRB/GW association. In Sec.3 we focus on the particular case of GW150914/GBM transient 150914. We summarize our results with some discussions in Sec.4.

\section{Some general implications of the GRB/GW association}
The GRB/GW association, if established, has some far-reaching implications, including for instance:
\begin{itemize}
\item {{\it A test of the merger origin of the ``old" or too far SGRBs/lsGRBs}}: the neutron star merger model for SGRBs/lsGRBs has been supported by host galaxy and afterglow observational data and in particular by the macronovae/kilonovae identified in SGRB 130603B, lsGRB 060614 and SGRB 050709. Nevertheless, these observational evidence are {\it indirect.} The GW signal associated with some SGRBs/lsGRBs, if detected in the future, will provide the direct evidence for neutron star merger scenario of these specific events. The comparison of these ``new $\gamma-$ray events" with the previous SGRBs/lsGRBs can serve as a valid test of the merger origin of these (old) events without GW data. If these new $\gamma-$ray events with an accompanying advanced-LIGO/Virgo GW signal are found to be similar to the (old) events without GW data in many aspects, the merger scenario for SGRBs/lsGRBs will be strongly supported (the same also holds for the events in the era of advanced LIGO/Virgo but beyond the sensitivity range of GW detectors). This implication, though looks to be apparently, is non-trivial in view of the relatively low detection rate of the GRB/GW association in the full-performance stage of advanced LIGO/Virgo (i.e., ${\cal R}_{\rm GRB/GW}\sim 1~{\rm yr^{-1}}$, which is much smaller than the SGRB/lsGRB detection rate that is $\sim 40$ per year for Fermi-GBM), with which the sample of GRB/GW association is expected to be still small in the next decade and the universal connection between SGRBs/lsGRBs and mergers can not be directly established.

\item {\it Constraining the mass of the accretion disk of the GRB and revealing the energy extraction process of the central engine}:
%The energy extraction process as well as the physical origin of the prompt emission of GRBs are still in debate \citep[e.g.,][]{Kumar2015}. In general the huge amount of energy needed to power the GRBs could be extracted either from neutrino processes (Consequently, the prompt emission is related to the photospheric radiation and subsequent internal shock emission) or magnetic processes (therefore the prompt emission is powered by the significant magnetic energy dissipation). But in reality it is very challenging to estimate the mass of the accretion disk ($M_{\rm disk}$) of %the GRB since usually
The energy output of the GRB central engine (an accretion disk $+$ central black hole system) depends on $M_{\rm BH}$, the accretion rate $(\dot{M})$, the spin of the black hole ($a$) and possibly also the structure of the disk. With the electromagnetic observational data the energy output of the central engine can be reasonably inferred, which however is not sufficient to break the degeneracies among parameters of $(M_{\rm BH}, ~\dot{M},~a)$, as stressed in \citet{Fan2011}. Therefore without additional assumption it is not possible to estimate the accretion disk mass ($M_{\rm disk}$) with the electromagnetic data alone. Fortunately, the situation for neutron star merger-driven GRBs could be much better. For some relatively ``nearby" SGRBs/lsGRBs with high quality gravitational wave data, the masses of the binary stars (and sometimes even the mass of the formed accretion disk) can be inferred \citep{Kiuchi2010}, with which $M_{\rm BH}$ and $a$ of the newly formed black hole can be reasonably evaluated \citep{Lee2000}. We can thus estimate $\dot{M}$ and $M_{\rm disk}$ in neutrino model and in the magnetic process model, respectively (see Sec.\ref{sec:Mdisk} for an illustration). If the GW data itself has been able to yield a reliable $M_{\rm disk}$, we can compare it with the estimated one and then distinguish between the energy extraction process. Otherwise if the $M_{\rm disk}$ found in a given model is significantly more massive than $\sim 0.1M_\odot$ (the upper limit of $M_{\rm disk}$ found in current numerical simulations), it is reasonable to rule out such a scenario.

\item {\it Directly measuring the velocity of the Gravitational wave}: In general relativity the velocity of gravitational wave ($v_{\rm g}$) is the speed of light ($c$). However, various gravity theories have been proposed in the literature and the GW velocity can be different from $c$ \citep[see][and the references therein]{Will1998}. The subluminal movement of gravitons has been extremely-tightly constrained (i.e., $\varsigma \equiv (c-v_{\rm g})/c<2\times 10^{-15}$) by the absence of gravitational Cherenkov radiation of the ultra-high energy cosmic rays detected on the earth \citep{Moore2001}. However, in the case of superluminal movement (i.e., $v_{\rm g}>c$),  currently the constraint is still ``loose", i.e., $(v_{\rm g}-c)/c<4\times 10^{-3}$ \citep{Baskaran2008}. The GRB/GW association, if established, can directly improve the constraint on the superluminal movement by many orders of magnitude.
\end{itemize}

\section{Implications of the GRB/GW association: the case of GW150914/GBM transient 150914}
%In this section we discuss the implications of the GRB/GW association in the tentative case of GW150914/GBM transient 150914. The following approaches can be directly applied to future GRB/GW events.

\subsection{Is GBM transient 150914 different from other SGRBs?}\label{sec:Comprision}

\begin{figure}
\begin{center}
\includegraphics[scale=0.45]{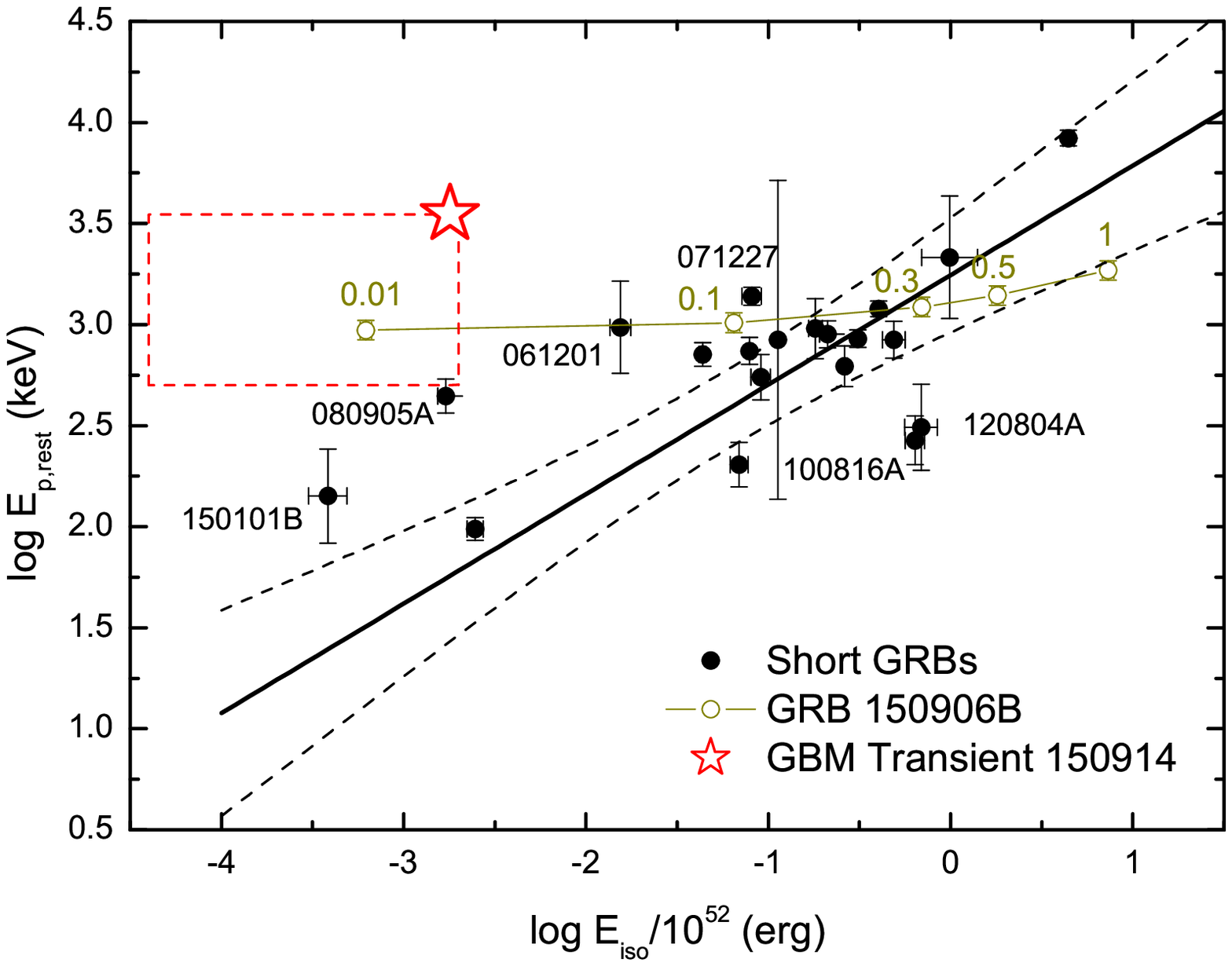}
\includegraphics[scale=0.45]{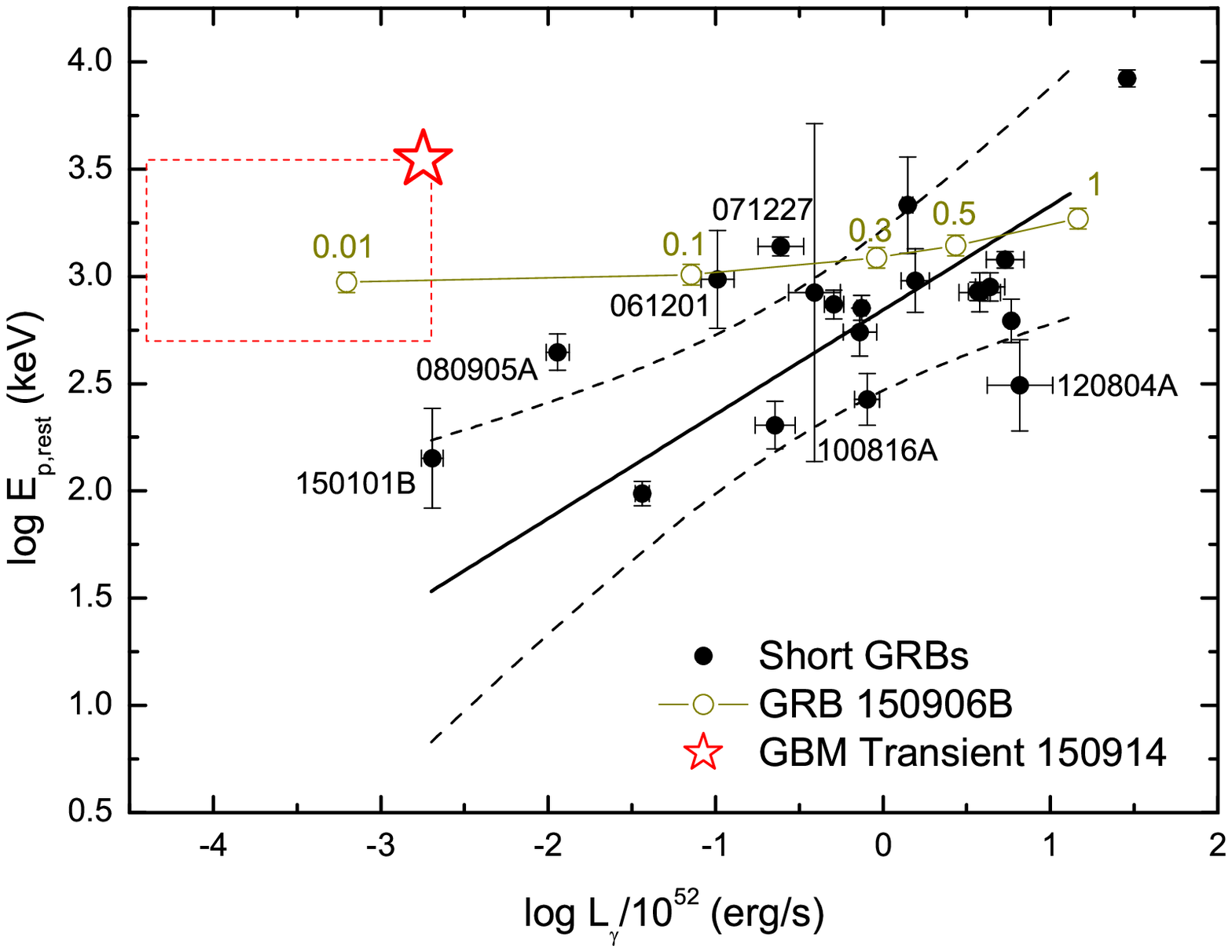}
\end{center}
\caption{The Upper and Lower panels are for the ``correlation" between the rest frame peak energy
$E_{\rm p,rest}$ and the isotropic total energy $E_{\rm iso}$ and the luminosity $L_{\rm \gamma}$ of SGRBs, respectively.
The filled circles represent the short GRBs with measured redshifts and spectral parameters updated up to Jan 1, 2016, the open circles represent GRB 150906B at different redshifts \citep[see also][]{ZhangF2015}, and the red pentagram represents GBM transient 150914. The red dashed rectangle represents the possible distribution of spectral peak energy and isotropic energy/luminosity for GBM Transient 150914. The solid and dashed lines are adopted from Fig.8 and Fig.9 of \citet{ZhangF2012ApJ}, which mark the allowed regions inferred from these early data. Some data are taken from \citet{ZhangF2012ApJ,ZhangF2015}, \citet{Gruber2012} and \citet{Gruber2014}, and some are analyzed in this work.
}
\label{fig:epeiso}
\end{figure}

A SGRB nature of the transient 150914 is favored in the Fermi GBM data analysis \citep[][see however Savchenko et al. (2016)]{Connaughton2016}. If indeed associated with GW150914, the luminosity $L_{\rm \gamma}=1.8^{+1.5}_{-1.0}\times 10^{49}~{\rm erg~s^{-1}}$ is in the low end of the distribution (with a duration of $\sim 1$ s we have $E_{\rm iso}\sim 2\times 10^{49}$ erg) while the spectral peak energy $E_{\rm peak}\sim 3$ MeV, however, is very high (Note that a Comptonized spectrum model yields $E_{\rm peak}\sim 3.5^{+2.3}_{-1.1}$ MeV and  the single power-law spectrum fit to the data up to the energy $\sim 4$ MeV gives an index of $-1.4^{+0.18}_{-0.24}$). As already noticed in \citet{Ruffini2015} and \citet{ZhangF2015}, the previous statistics of SGRBs \citep[e.g.,][]{ZhangF2012ApJ,Tsutsui2013,DAvanzo2014,Berger2014} found a typical $E_{\rm iso}\sim 10^{51}$ erg and $L_{\gamma}\sim 10^{52}~{\rm erg~s^{-1}}$ for $E_{\rm p,rest}=(1+z)E_{\rm peak}\sim 1$ MeV. Then the relatively low $L_\gamma$ and $E_{\rm iso}$ of the GBM transient 150914 likely renders it to be a distinguished outlier. To better check whether it is indeed the case, we have updated our previous analysis (i.e., Zhang et al. 2012) with a significantly extended sample of SGRBs with well measured $E_{\rm peak}$ and redshift ($z$). Our new $E_{\rm p,rest}-E_{\rm iso}$ and $E_{\rm p,rest}-L_{\gamma}$ diagrams are in Fig.\ref{fig:epeiso}, where a possible nearby event GRB 150906B \citep{Golenetskii2015,Levan2015} is also included. Interestingly we found that the current diagrams are not well consistent with the tight-correlations of $E_{\rm p,rest}-E_{\rm iso}$ and $E_{\rm p,rest}-L_{\gamma}$ reported in for example \citet[][i.e., see the previous allowed-regions marked by dashed lines in Fig.\ref{fig:epeiso}]{ZhangF2012ApJ}. In particular, there seems to be a new sub-group of low $L_\gamma$ ($E_{\rm iso}$) but high $E_{\rm p,rest}$ SGRBs \footnote{Indeed this possibility may be favored over the previous one since the chance to detect the first burst of a brand-new population in coincidence with the first GW event should be tiny.}, such as GRB 080905A \citep{Gruber2012}, GRB 150906B (if indeed at a distance of $\sim 52$ Mpc to the Galaxy \citep{Ruffini2015,ZhangF2015}) and the GBM transient 150914. Among our current sample GRB 090510 has the highest $E_{\rm p,rest} \sim 8.4$ MeV. Thanks to the very dense prompt emission, GRB 090510 is still marginally consistent with the $E_{\rm p,rest}-E_{\rm iso}$ and $E_{\rm p,rest}-L_{\gamma}$ correlations \citep[e.g.,][]{ZhangF2012ApJ,Tsutsui2013,DAvanzo2014}. The GBM transient 150914 may have the second highest $E_{\rm p,rest}$ but its $E_{\rm iso}$ and $L_\gamma$ are in the low end of the distribution, rendering such a source the most outstanding outlier of the $E_{\rm p,rest}-E_{\rm iso}$ and $E_{\rm p,rest}-L_{\gamma}$ correlations (Even if GRB 150906B is at $z=0.01$, GBM transient 150914 is a more distinct outlier).

{There are however some cautions. The location of GW150914 is poorly constrained, for all 11 positions along the LIGO arc analyzed by \citet[]{Connaughton2016}, a power-law is adequate to fit the spectrum of the transient. The $E_{\rm peak}$ reported in \citet[]{Connaughton2016} is from the Comptonized model fit assuming a source position at the northeastern tip of the southern lobe. Such a fit is not statistically preferred over the power-law and hence $E_{\rm peak}$ is uncertain. \citet[]{Savchenko2016} analyzed the data of INTEGRAL/SPI-ACS and
%presented an upper limit on the isotropic equivalent energy of $E_{\gamma}< 2\times 10^{48}~{\rm erg} ~\left({F_{\gamma} \over 10^{-7}\rm erg ~\rm %cm^{-2}}\right)\left({D \over 410 ~{\rm Mpc}}\right)^{2}$ and
reported upper limits on the fluence at the time of the event ranging from $2\times10^{-8} {\rm erg} ~\rm cm^{-2}$ to $10^{-6} {\rm erg}~\rm cm^{-2}$ in the 75 keV$-$2 MeV energy range for GRB spectral models (assuming two standard hard and soft GRB spectra with parameters $\alpha=-0.5, ~\beta=-1.5, ~E_{\rm peak}=1000$ keV and $\alpha=-1.5,~ \beta=-2.5, E_{\rm peak}=500$ keV) and sky positions. Greiner et al. (2016) reanalyzed the GBM data with PGStat and suggested that the GBM transient 150914 may be not an astrophysical event and the spectrum (fluence) is likely softer (lower) in comparison with typical short-hard GRBs. The best-fit spectral indices for positions along the LIGO arc cover the range -1.93 to -1.5 (with large errors) and the fluence covers the range $8\times10^{-8} {\rm erg} ~\rm cm^{-2}$ to $2.7\times10^{-7} {\rm erg}~\rm cm^{-2}$ in the 10 keV$-$1000 keV energy range (see Table 1 of Greiner et al. 2016). Motivated by these results, we consider a soft spectrum with $E_{\rm peak}\sim 500~{\rm keV}$ and $E_{\rm iso}\sim 4\times 10^{47}~{\rm erg}$ as the low end of the possible distribution. As shown in Fig. 1, an transient with such parameters may still be ``atypical" in the diagrams unless $E_{\rm peak}\leq 100$ keV.}

GBM transient 150914, if indeed associated with GW150914, has a binary black hole merger origin, different from other SGRBs that are believed to be powered by either double neutron star mergers or black hole-neutron star mergers. Therefore the dissimilarities in the prompt emission may reflect the different underlying physical processes. The other non-trivial possibility is that there is a group of SGRBs with low $L_\gamma$ and $E_{\rm iso}$ but high $E_{\rm p,rest}$ that are hard to detect unless take place ``nearby" (i.e., $z<0.1$). The nearby GRBs are rare in number, accounting for the rarity of such a group of ``emerging" events. So far, GBM transient 150914 is the unique candidate from double black hole merger. lsGRB 060614 and SGRB 050709 likely had a black hole-neutron star merger origin \citep{Yang2015,Jin2016}. For the rest SGRBs/lsGRBs, the progenitor stars are unknown and statistical studies in different kinds of mergers are not possible. In next decade when a reasonably large sample of GRBs with known origin is available, a statistical study of the prompt emission properties in different merger scenarios may better reveal the physical processes powering gamma-ray transients.

%%%%%%%%%%%%%%%%%%%%%%Fig.1%%%%%%%%%%%%%%%%%%%%%%%%%%

\begin{figure}
\begin{center}
\includegraphics[scale=0.6]{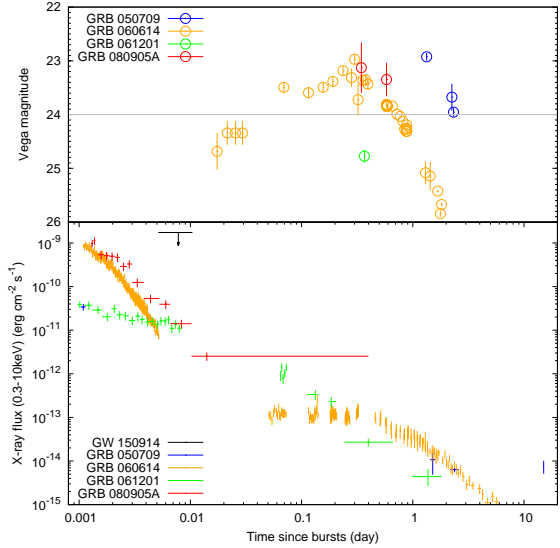}
\end{center}
\caption{The ``expected" afterglow emission of GBM transient 150914, which are ``generated" from the R-band (upper panel) and X-ray (lower panel) afterglow emission of several nearby SGRBs. The modifications include the corrections of fluxes due to the distance and $z$ shifts and the factor of $\sim 2\times 10^{49}~{\rm erg}/E_{\rm iso,i}$ to roughly correct the difference arising from different $E_{\rm iso}$ (according to the afterglow model \citep{Piran2004,Kumar2015}), where the subscript $i$ represents a given GRB presented in the figure. The X-ray and optical afterglow data are taken from \citet{Fong2015}. The 3 $\sigma$ upper limit of X-ray emission following GW150914 \citep{Serino2016} is also marked.
}
\label{fig:afterglow}
\end{figure}

After the GRB there should be relatively long-lasting afterglow emission. Instead of numerically estimating the forward shock afterglow, we ``generate" the expected emission with some nearby SGRBs, i.e., we collected the data of several nearby GRBs and converted them to the distance and roughly also the $E_{\rm iso}$ of GBM transient 150914 to get an ``overview" of the expected afterglow brightness (please see Fig.\ref{fig:afterglow}). For optical telescopes with a sensitivity of $\sim 24$th mag, the optical afterglow of GBM transient 150914 might be detectable within $\sim 1$ day after the burst. Due to the lack of wide-field sensitive X-ray monitor, with the very large location error, the detection of the forward shock X-ray afterglow emission is challenging. The prospect could be enhanced if there were X-ray flares, as observed in other GRB afterglows. The searches for optical and X-ray emission following GW150914 yielded null results, partly due to the inaccurate location  \citep[e.g.,][]{Smartt2016,Serino2016}.

%%%%%%%%%%%%%%%%%%%%%%Fig.2%%%%%%%%%%%%%%%%%%%%%%%%%%

\subsection{The mass of the accretion disk launching the outflow of GBM transient 150914}\label{sec:Mdisk}
A SGRB-like electromagnetic signal from a stellar-mass black hole binary merger is unexpected, as noticed in \citet{Connaughton2016}. A speculative scenario is the following: These two $\sim 30~M_\odot$ black holes had ``massive" disks. Some disk material survived in the merger and accreted onto the nascent $\sim 60~M_\odot$ black hole in a few seconds. Hence ultra-relativistic outflow was launched and the subsequent energy dissipation produced soft gamma-ray emission, as in the case of normal GRBs \citep[e.g.,][]{Piran2004,Kumar2015}. The other more speculative scenario is the reconnection of the magnetic fields confined in the two colliding disks. Alternative astrophysical scenarios giving rise to GW150914/GBM transient 150914 association can be found in the literature \citep[e.g.,][]{Loeb2016,Perna2016}. Instead of figuring out a detailed physical model of the prompt emission, below we estimate the mass of the accretion disk launching the outflow of GBM event 150914 (In this work we do not discuss the charged black hole model and refer the readers to Zhang (2016) and \citet{Savchenko2016}).

For the brief high energy transients, like GRBs, it is rather hard to estimate $M_{\rm disk}$ with the electromagnetic data alone since that the energy output of the central engine depends on $(M_{\rm BH},~\dot{M},~a)$ while the electromagnetic observational data alone can not break the degeneracies among these three parameters. For double neutron star mergers, the parameters of $M_{\rm BH}$ and $a$ can be relatively reasonably speculated, with which $\dot{M}$ and hence $M_{\rm disk}$ can be inferred \citep{Fan2011,LiuT2015}. Nevertheless, these earlier approaches are based on the ``hypothesized" $M_{\rm BH}$ and $a$. For GBM transient 150914, such approximations are not needed any longer.
With the gravitational wave data, the newly-formed black hole of GW150914 is found to have a mass $M_{\rm BH}\sim 62~M_\odot$ and a spin $a\sim 0.67$. Below we discuss the process(es) launching the outflow and then estimate $M_{\rm disk}$.

In general there are two kinds of physical processes that may launch ultra-relativistic energetic outflows. One invokes the neutrino/anti-neutrino annihilation (i.e., $\nu\bar{\nu}\rightarrow e^{+}e^{-}$; \citet{Eichler1989,Ruffert1998}). The other is the magnetic processes, for example the Blandford \& Znajek (1977) mechanism. We adopt an empirical relation of the neutrino/anti-neutrino annihilation luminosity proposed by  \citet{Zalamea10}, for $a=0.67$ which gives
%\begin{eqnarray} L_{\nu\bar{\nu}} &\approx & 10^{52}~{\rm erg~s^{-1}}x_{\rm ms}^{-4.8}({M_{_{\rm BH}}\over 3M_\odot})^{-3/2}\nonumber\\
%&&\left\{%
%\begin{array}{ll}
%    0, & \hbox{for $\dot{M}<M_{\rm ign}$;} \\
%    \dot{m}^{9/4}, & \hbox{for $M_{\rm ign}<\dot{M}<\dot{M}_{\rm trap}$;} \\
%    \dot{m}_{\rm trap}^{9/4}, & \hbox{for $\dot{M}\geq \dot{M}_{\rm trap}$,} \\
%\end{array}%
%\right.
%\label{eq:main}
%\end{eqnarray}
%where the accretion rate
%  $\dot{m}=\dot{M}/M_\odot~{\rm s^{-1}}$, $x_{\rm ms}=r_{\rm ms}(a)/r_{\rm g}$, $M_{\rm ign}=K_{\rm ign}(\alpha/0.1)^{5/3}$, ${M}_{\rm trap}=K_{\rm trap}(\alpha/0.1)^{1/3}$, and $\alpha$ is the viscosity. For $a=0~(0.95)$, it is found that $K_{\rm ign}=0.071~(0.021)~{\rm M_\odot~s^{-1}}$ and $K_{\rm trap}=9.3~(1.8)~{\rm M_\odot~s^{-1}}$, respectively \citep{Chen07}. The last stable orbit radius $r_{\rm ms}$ is given by \citep{Bardeen70ApJ}
%\begin{equation}
%r_{\rm ms}=r_{\rm g}\{3+Z_{2}\mp [(3-Z_{1})(3+Z_{1}+2Z_{2})]^{1/2}\}/2,
%\end{equation}
%where $r_{\rm g}=2GM_{\rm BH}/c^{2}$ ($G$ is Newton's constant and $c$ is the speed of light),
%$Z_{1}=1+(1-a^{2})^{1/3}[(1+a)^{1/3}+(1-a)^{1/3}]$ and $Z_{2}=(3a^{2}+Z_{1}^{2})^{1/2}$.
%For $a=0$ we have
%$r_{\rm ms}=3r_{\rm g}$, while for $a=1$ we have $r_{\rm ms}=r_{\rm g}/2$ or $r_{\rm ms}=9r_{\rm g}/2$ (retrograde).
%The retrograde case is irrelevant to the case of accretion disk and will not be discussed in this work any longer.
%For $a\sim 0.67$, we have $x_{\rm ms}\sim 1.53$ and (for $M_{\rm ign}<\dot{M}<\dot{M}_{\rm trap}$)
\begin{equation}
L_{\nu\bar{\nu}} \approx  1.4 \times 10^{49}~{\rm erg~s^{-1}} \dot{m}^{9/4}
%({x_{\rm ms}\over 1.53})^{-4.8}
({M_{\rm BH} \over 62M_\odot})^{-3/2},
\end{equation}
where the accretion rate is defined as $\dot{m}=\dot{M}/M_\odot~{\rm s^{-1}}$.
To account for the observed luminosity $L_{\gamma} \sim 4\times 10^{47}-2\times 10^{49}~{\rm erg~s^{-1}}$  of GBM transient 150914, we need $\dot{m}\sim 0.2-1.2$, which is too high to be realistic. If the outflow of GBM transient 150914 is highly collimated with an opening angle of $\theta_{\rm j}\sim 0.1$, we have $\dot{m}\sim 0.1~(L_\gamma/10^{49}~{\rm erg~s^{-1}})^{4/9}(\theta_{\rm j}/0.1)^{8/9}$ and hence an accretion disk mass
\[M_{\rm disk,\nu\bar{\nu}}\sim 0.1~(L_\gamma/10^{49}~{\rm erg~s^{-1}})^{4/9}(\theta_{\rm j}/0.1)^{8/9}~M_\odot,\]
which seems still be too high to be reasonable. We conclude that the neutrino/anti-neutrino annihilation process is disfavored.

The magnetic processes are known to be more efficient in launching relativistic outflow from hyper-accreting black holes \citep[e.g.,][and the references therein]{Fan2005ApJL,LiuT2015} and hence may be favored for the current event. In Blandford \& Znajek (1977) mechanism, the outflow luminosity is estimated to be \cite[see also][]{Lee2000}
\begin{equation}
L_{\rm BZ}\approx 4\times 10^{47} (a/0.67)^{2} (\dot{m}/10^{-4})~{\rm ergs~s^{-1}}.
\end{equation}
If collimated into an half-opening angle of $\theta_{\rm j}\sim 0.1$, the observed luminosity will be $L_{\gamma} \sim 2L_{\rm BZ}/\theta_{\rm j}^2 \sim 10^{50}~(a/0.67)^{2} (\dot{m}/10^{-4})(\theta_{\rm j}/0.1)^{-2}~{\rm erg~s^{-1}}$, which can account for the observation of GBM transient 150914 if %$\dot{m}\sim 4\times 10^{-7}-2\times 10^{-5}$. Correspondingly, the accretion disk should have a mass
\[M_{\rm disk,BZ} \sim 10^{-5}(L_{\gamma}/10^{49}~{\rm erg~s^{-1}})~M_\odot.\]
Such a massive transient accretion disk may suggest that the binary black holes were in dense medium. For example,
the double black hole binary system could be formed in a short distance capture (i.e., a black hole-star binary captures the other black hole)
and the dense medium was ejected from the star when the black holes mergered (Piran 2016, private communication; see also the talk at
https://gw150914.aei.mpg.de/program/tsvi-pirans-talk).
The material fallback from the collapse when the black hole formed can produce massive disks, too
(Katz 2016, private communication). However, the fallback accretion is not expected to last very long time.
Hence the merger should take place in a short time, which might be possible in some specific scenarios
\citep[e.g.,][]{Loeb2016,Perna2016}. {As for the specific single star model \citep{Loeb2016}, the challenge is how to give rise to a $\delta t$ as short as 0.4 s \citep{Woosley2016}.}
%Specific scenarios for forming massive disks surrounding the binary black holes can be found in \citet{Loeb2016} and \citet{Perna2016}.

Finally, we would like to point out that $\delta t \sim 0.4~{\rm s}$ and $T_\gamma \sim 1~{\rm s}$ are indeed consistent with that expected in the scenario of ``prompt" black hole formation $+$ subsequent magnetic jet launching and energy dissipation for SGRBs \citep[see Tab.1 of][]{LiX2016}.

%%%%%%%%%%%%%%%%%%%%%%%%%%%%%%%%%%%%%%%%%%%%%%%%

\subsection{Measuring gravitational wave velocity and constraining the graviton mass}\label{sec:GWV}
In general relativity theory, the speed of gravitational wave is the same as $c$. In other theories, the speed of gravitational wave however can differ from $c$ and one interesting possibility is that the gravitation were propagated by
a massive field. The non-zero graviton mass induces a modified
gravitational-wave dispersion relation and hence a modified group velocity that can be parameterized as \citep[e.g.,][]{Will1998}
$v_{\rm g}^{2}=(1-{m_{\rm g}^{2}c^{4}/E^2})c^{2}$,
where $m_{\rm g}$ and $E$ are the graviton rest mass and energy (usually associated
to its frequency via the quantum mechanical relation $E=hf$, where $h$  is Planck's constant and $f$ is the frequency), respectively.
In general we define the parameter $\varsigma\equiv (c-v_{\rm g})/c$ and a bound can be set by \citep[e.g.,][]{Will1998,Nishizawa2014,LiX2016}
\begin{equation}
|\varsigma|\leq 10^{-17}~\left({410 ~{\rm Mpc} \over D}\right)\left({\delta t\over 0.4~{\rm s}}\right).
\label{eq:constraint-2}
\end{equation}
Previously, limits on the speed of gravitational waves had been set indirectly in several model-dependent ways.
The solar system bound on the graviton mass yields a
$|\varsigma|\leq 10^{-8}$ \citep{Larson2000} and the bounds from pulsar timing is $|\varsigma|\leq 4\times 10^{-3}$ \citep{Baskaran2008}.
If the gravitational wave velocity is subluminal, then cosmic rays lose
their energy via gravitational Cherenkov radiation and cannot
reach the Earth. The observed ultra-high energy cosmic rays having an extragalactic or a galactic origin
suggests a $|\varsigma|\leq 2\times 10^{-19}$ or ~$\leq 2\times 10^{-15}$, respectively \citep{Caves1980,Moore2001}. Clearly our direct constraint on $|\varsigma|$ is much tighter than the solar system or the Galactic constraints. The full performance of advanced LIGO/Virgo network in 2020s is expected to be able to improve the constraint on $|\varsigma|$ by a factor of $\sim 100$, which can be comparable with the bound set by the extragalactic ultra-high energy cosmic rays.

The corresponding constraint on the mass of graviton is
\begin{equation}
m_{\rm g}\leq 8\times 10^{-22}~{\rm eV}~(|\varsigma|/10^{-17})^{1/2}(f/50~{\rm Hz}),
\end{equation}
and the bound on graviton
Compton wavelength $\lambda_{\rm g}=h/m_{\rm g}c$ is
\begin{equation}
\lambda_{\rm g}\geq 2\times 10^{17}~{\rm cm}.
\end{equation}
Comparing with the bounds summarized in Table 1 of \citet{Goldhaber2010}, our constraints on $m_{\rm g}$ and $\lambda_{\rm g}$ are weaker than some specific evaluation.

\section{Discussion and Conclusion}
Due to the (expected) low detection rate of GRB/GW association in the full-performance stage of advanced LIGO/Virgo, it was widely believed that the GRB/GW association will not be reliably established until 2020.  The merger-driven GRBs and their associated GW radiation, if both successfully detected, have some far-reaching implications, including for instance: (i) Testing the merger origin of the ``old" or too far short and long-short GRBs via the comparison of the physical properties of the events with and without GW detection; (ii) Constraining the mass of the accretion disk of the GRB and then revealing the energy extraction process of the central engine; (iii) Measuring the gravitational wave velocity directly/accurately.

On September 14, 2015 the two detectors of LIGO simultaneously detected a transient gravitational-wave signal GW150914 from the merger of a pair of $\sim 30 M_\odot$ black holes \citep{Abbott2016}. Usually a double black hole merger is unexpected to give rise to gamma-ray transient. The Fermi GBM observations, surprisingly, found a weak SGRB-like transient and the time/location coincidences favor the association between GW150904 and GBM transient 150914 \citep{Connaughton2016}. If correct, this would be the first time to identify a SGRB originated from a double black hole merger and suggest that the merger of much more massive black hole binaries may give rise to high energy transients that can serve as the electromagnetic counterparts of the gravitational wave signals.

We have compared GBM transient 150914 with other SGRBs with known redshift and well measured $E_{\rm peak}$ and found that such an event {may be} a distinct outlier in the $E_{\rm p,rest}-E_{\rm iso}$ and $E_{\rm p,rest}-L_{\gamma}$ diagrams (see Fig.\ref{fig:epeiso}). The dissimilarities of GBM transient 150914 with other SGRBs might be attributed to its specific binary-black-hole merger origin. However, together with GRB 080905A and possibly also GRB 150906B (if indeed very nearby with a $z\sim 0.01$), there might be a ``new" group of SGRBs with low $L_\gamma$ and $E_{\rm iso}$ but high $E_{\rm p,rest}$ that are hard to detect unless they took place ``nearby". With the current limited sample of (nearby) SGRBs, it is hard to conclude wether the ``peculiarity" of prompt emission of GBM transient 150914 is ``intrinsic" or not (see Sec.\ref{sec:Comprision}).

The physical origin of GBM transient 150914 is unclear. A speculative process is the hyper accretion of the disk material survived in the merger onto the nascent black hole. Within such a scenario we show that the outflow powering GBM transient 150914 was likely launched via some magnetic progresses. The mass of the newly-formed black hole as well as its spin parameter inferred from the gravitational wave data \citep{Abbott2016} provide the first chance to evaluate the accretion rate/accretion disk mass without making additional assumptions on the needed physical parameters. The estimated accretion disk mass is $\sim 10^{-5}(L_{\gamma}/10^{49}~{\rm erg~s^{-1}})~M_\odot$, implying that the binary black hole progenitors were in dense medium (see Sec.\ref{sec:Mdisk}).

{If confirmed, the association between GBM transient 150914 and GW150914 would also provide the first opportunity to directly measure the velocity of the gravitational wave and the difference between the gravitational wave velocity and the speed of the light should be within a factor of $10^{-17}$} (see eq.(\ref{eq:constraint-2}) in Sec.\ref{sec:GWV}; see also Ellis et al. 2016), which is nicely in agreement with the prediction of the general relativity. With the successful performance of advanced LIGO/Virgo network in 2020s, the bound on $|\varsigma|$ is expected to be tightened by a factor of $\sim 100$.

Finally we would like to point out that though we focus on the implications of the GRB/GW association in the tentative case of GW150914/GBM transient 150914, the approaches are general and can be directly applied to future GRB/GW events.

\section*{Acknowledgments}
We thank the anonymous referee, J. Katz and S. Desai for suggestions/discussions. This work was supported in part by National Basic Research Programme of China (No. 2013CB837000 and No. 2014CB845800), NSFC under grants No. 11525313 (i.e., the Funds for Distinguished Young Scholars), No. 11433009, No. U1331101, No. 11273063 and No. 11163003, and the Strategic Priority Research Program (Grant No. XDB09000000). This work was also supported by the Joint NSFC-ISF Research Program, jointly funded by the National Natural Science Foundation of China and the Israel Science Foundation (No. 11361140349). F.-W.Z. also acknowledges the support by the Guangxi Natural Science Foundation (No. 2013GXNSFAA019002) and the project of outstanding young teachers' training in higher education institutions of Guangxi.
\\

\clearpage


\begin{thebibliography}{}

\bibitem[Abbott et al. (2016a)]{Aasi2013} Abbott, B. P., et al. (LIGO Scientific Collaboration, Virgo Collaboration),
2016a, Living Rev. Relativity, 19, 1

\bibitem[Abbott et al. (2016b)]{Abbott2016} Abbott, B. P. et al. (LIGO Scientific Collaboration, Virgo Collaboration), 2016b, Phys. Rev. Lett., 116, 061102

%\bibitem[Abadie et al. (2010)]{Abadie2010} Abadie, J., Abadie, J., Abbott, B. P., et al. 2010, CQGra, 27, 173001

\bibitem[Baskaran et al. (2008)]{Baskaran2008}  Baskaran, D., Polnarev,  A. G., Pshirkov, M. S., and Postnov, K. A.
%¡°Limits on the Speed of Gravitational Waves from Pulsar Timing,¡±
2008, Phys. Rev. D 78, 044018.

%\bibitem[Bardeen et al. (1972)]{Bardeen70ApJ} Bardeen, J. M., Press, W. H., \& Teukolsky, S. A. 1972, ApJ, 178, 347

\bibitem[Barnes \& Kasen (2013)]{Barnes2013} Barnes, J. \& Kasen, D. 2013, ApJ, 773, 18.

\bibitem[Berger et al. (2013)]{Berger2013} Berger, E., Fong, W., \& Chornock, R. 2013, ApJL, 744, L23

\bibitem[Berger (2014)]{Berger2014} Berger, E., 2014, ARA\&A, 52, 43

\bibitem[Blandford \& Znajek (1977)]{Blandford1977} Blandford, R. D., \& Znajek, R. L. 1977, MNRAS, 179, 433

%\bibitem[Bufano et al. (2012)]{Bufano2012ApJ} Bufano, F., Pian, E., Sollerman, J., et al. 2012, ApJ, 753, 67

%\bibitem[Chen \& Beloborodov (2007)]{Chen07}Chen, W. X., \& Beloborodov, A. M. 2007, ApJ, 657, 383

\bibitem[Caves (1980)]{Caves1980} Caves, C. M. 1980, Ann. Phys., 125, 35

\bibitem[Clark \&  Eardley (1977)]{Clark1977} Clark, J. P. A., \& Eardley, D. M., 1977, ApJ, 215, 311

\bibitem[Clark et al. (2015)]{Clark2015} Clark, J. et al. 2015, ApJ, 809, 53

\bibitem[Connaughton et al. (2016)]{Connaughton2016} Connaughton, V., et al. 2016, ApJL in press (arXiv:1602.03920)

\bibitem[D'Avanzo et al.(2014)]{DAvanzo2014} D'Avanzo, P.,
Salvaterra, R., Bernardini, M.~G., et al.\ 2014, \mnras, 442, 2342

\bibitem[Della Valle et al. (2006)]{DellaValle2006} Della Valle, M., Chincarini, G., Panagia, N.,  et al. 2006, Natur, 444, 1050

\bibitem[Eichler et al. (1989)]{Eichler1989} Eichler D.,  Livio M.,  Piran T., \& Schramm D. N. 1989, Natur, 340, 126

\bibitem[Ellis et al. (2016)]{Ellis2016} Ellis, J. et al. 2016, arXiv:1602.04764

\bibitem[Evans et al. (2009)]{Evans2009} Evans, P. A., Beardmore, A. P., Page, K. L., et al. 2009, MNRAS, 397, 1177

\bibitem[Fan \& Wei (2011)]{Fan2011} Fan, Y. Z., \& Wei, D. M. 2011, ApJ, 739, 47

\bibitem[Fan et al. (2005)]{Fan2005ApJL} Fan, Y. Z., Zhang, B., \& Proga, D. 2005, ApJL, 635, L129

\bibitem[Finn \& Sutton (2002)]{Finn2002} Finn, L. S., \& Sutton, P. J., 2002,  Phys.
Rev. D, 65, 044022

%\bibitem[Fong et al. (2010)]{Fong2010} Fong, W., Berger, E., \& Fox, D. B. 2010, ApJ, 708, 9

%\bibitem[Fong \& Berger (2013)]{Fong2013} Fong, W., \& Berger, E., 2013, ApJ, 776, 18

\bibitem[Fong et al. (2015)]{Fong2015} Fong, W. F. et al.
%A Decade of Short-duration Gamma-Ray Burst Broadband Afterglows: Energetics, Circumburst Densities, and Jet Opening Angles.
2015, ApJ, 815, 102

\bibitem[Fynbo et al. (2006)]{Fynbo2006} Fynbo, J. P. U., Watson, D., Th\"one, C. C., et al. 2006, Natur, 444, 1047

\bibitem[Gal-Yam et al. (2006)]{Gal-Yam2006} Gal-Yam, A., Fox, D. B., Price, P. A., et al. 2006, Natur, 444, 1053

%\bibitem[Gao et al. (2015)]{Gao2015} Gao, H., Ding, X., Wu, X. F., Dai, Z. G. \& Zhang, B. 2015, ApJ, 807, 163

\bibitem[Gehrels et al. (2006)]{Gehrels2006} Gehrels, N., Norris, J. P., Barthelmy, S. D., et al. 2006, Natur, 444, 1044

%\bibitem[Gehrels et al. (2005)]{Gehrels2005} Gehrels, N., Sarazin, C. L., O'Brien, P. T., et al. 2005, Natur, 437, 851

\bibitem[Golenetskii et al. (2015)]{Golenetskii2015} Golenetskii, S. et al. 2015, GCN Circ. 18259 (http://gcn.gsfc.nasa.gov/gcn3/18259.gcn3)

\bibitem[Goldhaber \& Nieto (2010)]{Goldhaber2010} Goldhaber, A. S., \&  Martin Nieto, M. M. 2010, Rev. Mod. Phys., 82, 939

%\bibitem[Grossman et al. (2014)]{Grossman2014} Grossman, D., Korobkin, O., Rosswog, S., \& Piran, T. 2014, MNRAS, 439, 757

\bibitem[Greiner et al.(2016)]{Greiner2016} Greiner, J., Burgess, J.~M., Savchenko, V., \& Yu, H.-F.\ 2016, arXiv:1606.00314

\bibitem[Gruber (2012)]{Gruber2012} Gruber, D., 2012, Proceedings of Science PoS (GRB 2012), 007 (http://arxiv.org/abs/1207.4620)

\bibitem[Gruber et al. (2014)]{Gruber2014} Gruber, D., et al. 2014, ApJS, 211, 12

%\bibitem[Hotokezaka et al. (2013)]{Hotokezaka2013} Hotokezaka, K.,  Kyutoku, K., Tanaka, M., et al. 2013, ApJL, 778, L16

\bibitem[Jin et al. (2015)]{Jin2015} Jin, Z. P., et al. 2015, ApJL, 811, L22

\bibitem[Jin et al. (2016)]{Jin2016} Jin, Z. P., et al. 2016, Nat. Commun. submitted (arXiv:1603.07869)

\bibitem[Kasen et al. (2013)]{Kasen2013} Kasen, D., Badnell, N. R. \& Barnes, J. 2013, ApJ, 774, 25

\bibitem[Kiuchi et al. (2010)]{Kiuchi2010} Kiuchi, K., Sekiguchi, Y., Shibata, M., \& Taniguchi, K. 2010, Phys. Rev. Lett.,
104, 141101

\bibitem[Kouveliotou et al. (1993)]{Kouveliotou1993} Kouveliotou, C., C. A. Meegan, G. J. Fishman, N. P. Bhat, M.
S. Briggs, T. M. Koshut, W. S. Paciesas, and G. N. Pendleton,
1993, ApJL, 413, L101

%\bibitem[Korobkin et al. (2012)]{Korobkin12} Korobkin, O., Rosswog, S.,  Arcones, A., \& Winteler, C. 2012, MNRAS, 426, 1940

%\bibitem[Kulkarni (2005)]{Kulkarni2005} Kulkarni, S. R. 2005, arXiv:astro-ph/0510256

\bibitem[Kumar \& Zhang (2015)]{Kumar2015} Kumar, P., \& Zhang, B., 2015, PhR, 561, 1

%\bibitem[Kyutoku et al. (2015)]{Kyutoku2015} Kyutoku, K., Ioka, K., Okawa, H., Shibata, M., \& Taniguchi, K. 2015, arXiv:1502.05402

%\bibitem[Leibler et al. (2010)]{Leibler2010} Leibler C. N., \& Berger E. 2010, ApJ, 725, 1202

\bibitem[Lee et al. (2000)]{Lee2000} Lee, W. H., Wijers, R. A. M. J., \& Brown, G. E. 2000, Phys. Rep., 325, 83

\bibitem[Levan et al. (2015)]{Levan2015} Levan, A. J., Tanvir, N. R., \& Hjorth, J., 2015, GCN Circ. 18263 (http://gcn.gsfc.nasa.gov/gcn3/18263.gcn3)

\bibitem[Larson \& Hiscock (2000)]{Larson2000} Larson, S. L., \& Hiscock, W. A. 2000, Phys. Rev. D, 61, 104008

\bibitem[Li \& Paczy\'{n}ski (1998)]{Li1998} Li, L.-X., \& Paczy\'{n}ski, B. 1998, ApJL, 507, L59

\bibitem[Li et al. (2016)]{LiX2016} Li, X., Hu, Y. M., Fan, Y. Z., \& Wei, D. M. 2016, ApJ in press (arXiv:1601.00180)

\bibitem[Liu et al. (2015)]{LiuT2015} Liu, T., Lin, Y. Q., Hou, S. J., \&  Gu, W. M. 2015, ApJ, 806, 58

%\bibitem[Lippuner \& Roberts (2015)]{Lippuner2015} Lippuner, J. \& Roberts, L. F. 2015, arXiv:1508.03133

\bibitem[Loeb (2016)]{Loeb2016} Loeb, A., 2016, arXiv:1602.04735

%\bibitem[Mangano et al. (2007)]{Mangano2007} Mangano, V.,  Holland, S. T., Malesani, D. et al. 2007, A\&A, 470, 105

\bibitem[Metzger et al. (2010)]{Metzger2010} Metzger, B. D., Mart\'{i}nez-Pinedo, G., Darbha, S. et al. 2010, MNRAS, 406, 2650

%\bibitem[Metzger \& Fern\'{a}ndez (2014)]{Metzger2014} Metzger, B. D. \& Fern\'{a}ndez, R. 2014, MNRAS, 441, 3444

\bibitem[Moore \& Nelson (2001)]{Moore2001} Moore, G. D.  and  Nelson, A. E., 2001, JHEP, 0109, 023

%\bibitem[Narayan et al. (1992)]{Narayan1992} Narayan, R., Paczynski, B., \& Piran, T. 1992,  ApJL, 395, L83

\bibitem[Nishizawa \& Nakamura (2014)]{Nishizawa2014} Nishizawa, A., \& Nakamura, T., 2014, Phys. Rev. D, 90, 044048

%\bibitem[Olivares et al. (2012)]{Olivares2012} Olivares, E. F., Greiner, J., Schady, P. et al. 2012, A\&A, 539, A76

\bibitem[Perna et al. (2016)]{Perna2016} Perna, R., Lazzati, D., \& Giacomazzo, B., 2016, arXiv:1602.05140

\bibitem[Piran (2004)]{Piran2004} Piran, T., 2004, RvMP, 76, 1143

%\bibitem[Piran et al. (2014)]{Piran2014} Piran, T., Korobkin, O., Rosswog, S., 2014, arXiv:1401.2166

%\bibitem[Rosswog (2005)]{Rosswog2005} Rosswog, S. 2005, ApJ, 634, 1202

\bibitem[Ruffert \& Janka (1998)]{Ruffert1998} Ruffert, M., \& Janka, H. T. 1998, A\&A, 338, 535

\bibitem[Ruffini et al. (2015)]{Ruffini2015} Ruffini, R., et al. 2015, GCN Circ. 18296
(http://gcn.gsfc.nasa.gov/gcn3/18296.gcn3)

%\bibitem[Sari et al. (1999)]{Sari1999} Sari, R. Piran, T. \& Halpern, 1999, ApJL, 519, L17

\bibitem[Savchenko et al. (2016)]{Savchenko2016} Savchenko, V., et al. 2016, ApJL, submitted (arXiv:1602.04180)

%\bibitem[Schlafly \& Finkbeiner(2011)]{2011ApJ...737..103S} Schlafly, E.~F., \& Finkbeiner, D.~P.\ 2011, ApJ, 737, 103

\bibitem[Serino et al. (2016)]{Serino2016} Serino, M., et al. 2016, GCN Circ. 19013
(http://gcn.gsfc.nasa.gov/gcn3/19013.gcn3)

%\bibitem[Schlegel, Finkbeiner \& Davis(1998)]{Schlegel98} Schlegel, D. J., Finkbeiner, D. P., \& Davis, M.\ 1998, ApJ, 500, 525

\bibitem[Smartt et al. (2016)]{Smartt2016} Smartt, S. J. et al. 2016, arXiv:1602.04156

\bibitem[Smartt et al. (2016)]{Swift2016} Smartt, S. J. et al. 2016, arXiv:1602.04156

%\bibitem[Tanaka \& Hotokezaka (2013)]{Tanaka2013} Tanaka, M., \& Hotokezaka, K. 2013, ApJ, 775, 113

%\bibitem[Tanaka et al. (2014)]{Tanaka2014} Tanaka, M., Hotokezaka, K., Kyutoku, K. et al. 2014, ApJ, 780, 31

\bibitem[Tanvir et al. (2013)]{Tanvir2013} Tanvir, N. R., Levan, A. J., Fruchter, A. S. et al. 2013, Natur, 500, 547

\bibitem[Tsutsui et al.(2013)]{Tsutsui2013} Tsutsui, R., Yonetoku,
D., Nakamura, T., Takahashi, K., \& Morihara, Y.\ 2013, \mnras, 431, 1398

%\bibitem[Wanajo et al. (2014)]{Wanajo2014} Wanajo, S., Sekiguchi, Y., Nishimura, N. et al. 2014, ApJL, 789, L39

%\bibitem[Wanderman \& Piran (2015)]{Wanderman2014} Wanderman, D., \& Piran, T. 2015, MNRAS, 448, 3026

\bibitem[Will (1998)]{Will1998} Will, C. M., 1998, Phys. Rev. D., 57, 2061

\bibitem[Woosley(2016)]{Woosley2016} Woosley, S.~E.\ 2016, \apjl, 824, L10

%\bibitem[Xu et al. (2009)]{Xu2009} Xu, D., Starling, R. L. C., Fynbo, J. P. U. et al. 2009, ApJ, 696, 971

\bibitem[Yang et al. (2015)]{Yang2015} Yang, B., Jin, Z. P., Li, X. et al. 2015, Nat. Commun., 6, 7323

\bibitem[Zalamea \& Beloborodov (2011)]{Zalamea10} Zalamea, I., \& Beloborodov, A. M. 2011, MNRAS, 410, 2302

\bibitem[Zhang(2016)]{2016arXiv160204542Z} Zhang, B.\ 2016, arXiv:1602.04542

\bibitem[Zhang et al.(2007)]{Zhang2007} Zhang, B., Zhang, B. B., Liang, E. W. et al. 2007, ApJL, 655, L25

\bibitem[Zhang et al.(2012)]{ZhangF2012ApJ} Zhang, F.-W., Shao, L.,
Yan, J.-Z., \& Wei, D.-M.\ 2012, ApJ, 750, 88

\bibitem[Zhang et al.(2015)]{ZhangF2015} Zhang, F.-W., Zhang, B.,
\& Zhang, B.\ 2015, GCN Circ. 18298 (http://gcn.gsfc.nasa.gov/gcn3/18298.gcn3)

\end{thebibliography}
\end{document}